\documentclass[conference]{IEEEtran}
\IEEEoverridecommandlockouts
\usepackage{amsmath,amssymb,amsfonts}
\usepackage{algorithmic}
\usepackage{graphicx}
\usepackage{textcomp}
\usepackage{xcolor}
\usepackage[style=ieee,sorting=none]{biblatex}
\def\BibTeX{{\rm B\kern-.05em{\sc i\kern-.025em b}\kern-.08em
    T\kern-.1667em\lower.7ex\hbox{E}\kern-.125emX}}
\begin{document}

\title{Optimizing Temporal Waveform Analysis: A Novel Pipeline for Efficient Characterization of Left Coronary Artery Velocity Profiles\\
\thanks{Research reported in this publication was supported by the National Institute On Aging of the National Institutes of Health under Award Number DP1AG082343. The content is solely the responsibility of the authors and does not necessarily represent the official views of the National Institutes of Health.}
}

\author{\IEEEauthorblockN{Justen R. Geddes}
\IEEEauthorblockA{\textit{Biomedical Engineering} \\
\textit{Duke University}\\
Durham NC, USA \\
justen.geddes@duke.edu}
\and
\IEEEauthorblockN{Amanda Randles}
\IEEEauthorblockA{\textit{Biomedical Engineering} \\
\textit{Duke University}\\
Durham NC, USA \\
amanda.randles@duke.edu}
}
\maketitle

\begin{abstract}
Continuously measured arterial blood velocity can provide insight into physiological parameters and potential disease states. The efficient and effective description of the temporal profiles of arterial velocity is crucial for both clinical practice and research. We propose a pipeline to identify the minimum number of points of interest to adequately describe a velocity profile of the left coronary artery. This pipeline employs a novel operation that ``stretches" a baseline waveform to quantify the utility of a point in fitting other waveforms. Our study introduces a comprehensive pipeline specifically designed to identify the minimal yet crucial number of points needed to accurately represent the velocity profile of the left coronary artery.  Additionally, the only location-dependent portion of this pipeline is the first step, choosing all of the possible points of interest. Hence, this work is broadly applicable to other waveforms. This versatility paves the way for a novel non-frequency domain method that can enhance the analysis of physiological waveforms. Such advancements have potential implications in both research and clinical treatment of various diseases, underscoring the broader applicability and impact.

\end{abstract}

\begin{IEEEkeywords}
Coronary artery, velocity waveform
\end{IEEEkeywords}

%------------------ INTRODUCTION ------------------
\section{Introduction}
%Temporal velocity waveforms are widespread in clinical medicine, as they influence a wide variety of vital physiologic metrics such as cardiac output and endothelial wall shear stress. However, systematically quantifying a given temporal velocity profile remains a key challenge. In this study, we develop a pipeline to achieve this and apply it to simulated left coronary artery waveforms.

Temporal velocity waveforms play a critical role in clinical diagnostics, influencing key physiological metrics such as cardiac output and shear stress of the endothelial wall. However, systematic quantification of these waveforms poses a substantial challenge. This study aims to develop a method for efficient quantification and apply it to simulated left coronary artery waveforms. Despite previous studies partially quantifying coronary velocity profiles, there is still a notable absence of comprehensive methods to capture the temporal shape of these waveforms. Our pipeline, inspired by existing research, seeks to fill this gap by determining the minimum number of metrics essential to quantify the temporal shape of the velocity waveforms of the left coronary artery.

Temporal quantification of physiological waveforms is largely lacking, and coronary velocity profiles remain particularly understudied. Studies have partially quantified coronary waveforms in patients \cite{davies2006use,hozumi1998noninvasive,kajiya1993velocity,ofili1993coronary, schiemann2006mr} and simulated waveforms \cite{chidyagwai2022characterization,gutierrez2019hemodynamic,rizzini2020does}. Previous attempts to quantify coronary artery velocity have focused on coarse metrics such as the area under the curve for portions of the cycle and the maximum velocity \cite{hozumi1998noninvasive,ofili1993coronary}. The pipeline presented here is inspired by these studies as well as Fraser et al. \cite{fraser2008characterization}, who quantified the temporal shape of a velocity waveform in an abdominal aortic aneurysm using 14 points of interest automatically identified, but did not assess the importance of these points to reduce their number. There is a lapse in the literature for the minimum number of metrics needed to quantify the temporal shape of a left coronary velocity waveform. 

To remedy the lack of efficient quantification of temporal blood waveforms, we develop a pipeline that can identify the smallest number of points needed to quantify the temporal shape of a velocity waveform. We apply this pipeline to left coronary velocity waveforms as a proof of concept; however, this pipeline can be generalized to quantify other waveforms.

%------------------ METHODS ------------------
\section{Methods}
This pipeline begins by identifying all possible points of interest on the simulated waveforms. We then select the minimum number of points by defining a novel transformation of the waveforms and assessing a ``base" waveform's ability to replicate a ``target" waveform via this transformation. We argue that the points of interest that allow a base waveform to best match target waveforms are the most important for the characterization of the waveform.

\subsection{Coronary artery velocity waveforms}
Due to the scarcity of coronary artery ostium velocity data sets, we used 4.5 million simulated waveforms from \cite{tanade2023cloud}. These waveforms were created by scaling a template waveform from the literature to match the maximum velocity and the length of the cardiac cycle for a given constant stroke volume. Next, synthetic noise representing uncertainty from Doppler measurements of coronary velocity was added to both the velocity and time components of the waveform. We then take a representative subsample of 1,000 waveforms from this larger set while retaining the variation in cardiac cycle length, systolic flow, and diastolic flow.

% This method created a large database that we then sampled from while retaining the variation in time and velocity. To do so we first binned all of the waveforms based on the integral over systole, the integral over diastole, and cardiac cycle length. We then randomly select waveforms until we have 1,000 representative waveforms from the 4.5 million such that the relative frequency of each bin corresponds to the frequency of that bin in the larger set, e.g. if 1\% of the 4.5 million waveforms belong in bin $i$, then in our representative set 1\% of the 1,000 waveforms belong in bin $i$. 

\subsection{Identifying points of interest on the left coronary artery ostium waveform}
\begin{figure}[tbp]
% \centerline{\includegraphics[width=\columnwidth]{figures/points_on_wave.png}}
\centerline{\includegraphics[width=\columnwidth]{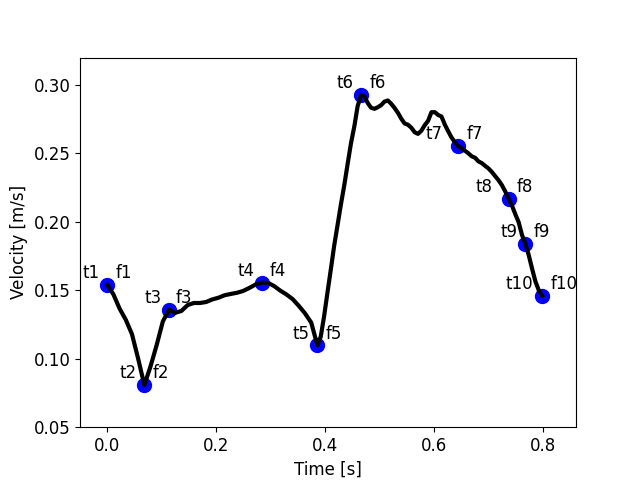}}
\caption{All 20 values labeled on an example waveform. Definitions of points can be found in Table \ref{tab: point defs}. Note that each point has a time and velocity component.}
\label{fig: all points on waveform}
\end{figure}

 As inspired by \cite{fraser2008characterization} we first select points of interest on the base waveform. To start, we select 10 points (20 values, one time value, and one velocity value for each point) that appear important to the shape of left coronary artery waveforms and hold physiological significance. At this step in the pipeline, we choose the maximum number of points that appear significant, and we reduce the number later. These points are presented in Table \ref{tab: point defs} and are shown in Fig. \ref{fig: all points on waveform}.

% These values produce a high-dimensional parameter space to explore and we therefore wish to use the minimal number of points possible that still encapsulates the variation in waveforms. To find the minimal number of needed points we first define a transformation that acts on waveforms.

\begin{table}[tbp]
    \centering
    \caption{Definition of point locations on waveforms. $f_i$ refers to the velocity of point $i$. }
    \begin{tabular}{|c l|}
    \hline
        Point number & Description  \\
        \hline
         1 & Beginning of systole\\
         2 & Systolic valley \\
         3 & Systolic shoulder\\
         4 & Systolic max \\
         5 & Beginning of diastole \\
         6 & Diastolic peak \\
         7 & 3/4 ($f_6 - f_{10}$) + $f_{10}$ \\
         8 & 1/2 ($f_6 - f_{10}$) + $f_{10}$\\
         9 & 1/4 ($f_6 - f_{10}$) + $f_{10}$ \\
         10 & End of cardiac cycle\\
         \hline
    \end{tabular}
    \label{tab: point defs}
\end{table}

\subsection{Novel stretching transformation}
To change the shape of a waveform we propose a novel transformation that can be conceptualized as ``stretching". If one simply changes the value of a point of interest that point will then appear to be discontinuous to the rest of the waveform. We assume that if one point, either time or velocity, increases/decreases then the surrounding points should also increase/decrease.

To stretch the waveform we derive a transformation inspired by linear splines. We take a ``base" waveform ($t_f$ \& $f$) and seek to stretch that waveform so it fits a ``target" waveform ($t_w$ \& $w$). Let $P_i$ denote the point of interest $i$ on the base waveform that is being matched, $\hat{P}_i$ be the point of interest $i$ on the target waveform, and $p_k$ denote the $k^{\text{th}}$ value of the base waveform vectors between the two points of interest $i-1$ and $i$.

% \begin{align*}
%     f &: \text{``Base" waveform velocity vector}\\
%     w &: \text{Target waveform velocity vector}\\
%     g &: \text{Waveform velocity vector computed} \\
%     & \hspace{7mm} \text{by stretching $f$}\\
%     t_f &: \text{``Base" waveform time vector}\\
%     t_w &: \text{Target waveform velocity time vector}\\
%     t_g &: \text{Waveform velocity time vector computed} \\
%     & \hspace{7mm} \text{ by stretching $t_f$}
% \end{align*}
% \begin{align*}
%     P_i &: \text{Point $i$ on the base waveform ($t_f$ \& $f$)}\\ 
%     &\hspace{5mm} \text{ that is being matched.} \\ 
%     \hat{P}_i &: \text{Point $i$ on the target waveform ($t_w$ \& $w$)}\\
%     p_k &: \text{Values on the base time or velocity vectors between}\\ &\hspace{5mm} \text{ the two points $i-1$ and $i$ defined above}
% \end{align*}

\begin{figure*}[tbp]
    \centering
    \includegraphics[width=\textwidth]{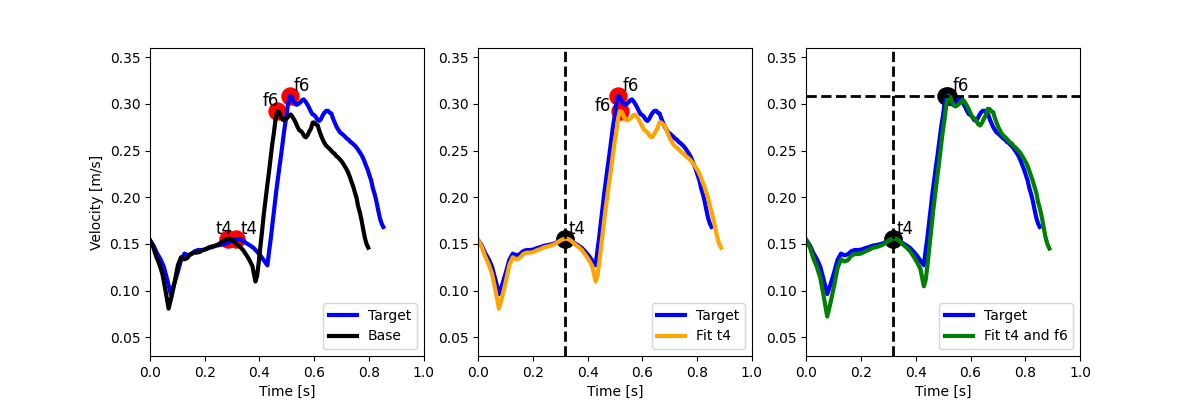}
    \caption{The process of ``stretching" waveforms. From left to right, the first panel shows the ``base" waveform (black) that will be stretched to match the target (blue) waveform. The second panel shows the result of stretching the waveform along the time axis to match the 4$^\text{th}$ time point (yellow). The third panel shows the result of stretching the waveform in the second panel to match the 6$^\text{th}$ velocity point (green).}
    \label{fig: stretching waveforms}
\end{figure*}

Note that $P_1$ is matched via addition (translational shift in time and velocity) so this recursive process has an adequate base case. Let $T: \mathbb{R}^n \rightarrow \mathbb{R}^n$ be the transformation from the base time or velocity vector to the new waveform and $T_k: \mathbb{R} \rightarrow \mathbb{R}$ be the operation for entry $k$ out of $n$ total entries. The transformation process for both time and velocity is identical. Thus, we define $T$ as a function of velocity vectors, but it equally applies to time vectors. We define 
\begin{align*}
    T_k(p_k) &= \underbrace{\frac{p_k - P_{i-1}}{P_i-P_{i-1}}}_{a_k} \underbrace{\left(\hat{P}_i-\hat{P}_{i-1}\right)}_{x_k} + \underbrace{\hat{P}_{i-1}}_{u_k},\\
     % &= \frac{\hat{P}_{i}-\hat{P}_{i-1}}{P_i-P_{i-1}} \left(p_k - P_{i-1}\right) + \hat{P}_{i-1},\\
     \rightarrow T(f) &= a \odot x + u, \hspace{1cm} a, x, u \in \mathbb{R}^n,
\end{align*}
\noindent where $\odot$ denotes the Hadamard product (element-wise multiplication). Note that $a$ need only be computed once per base waveform, while $x$ and $u$ need to be computed for each target waveform. An illustration of this stretching operation is depicted in Fig. \ref{fig: stretching waveforms}.

\subsection{Reducing the number of points of interest}
Examining all 20 identified points of interest would be a complex task. Furthermore, it would be unclear which points are more important than others. It is also likely that some points are redundant and do not hold insight when other points are used. To find the fewest points needed to characterize a waveform, we seek to find the smallest subset of points that allow an adequate recreation of the target waveforms using the stretching transformation. To assess which subset is best, we begin with a base waveform and fit that waveform to the 1,000 representative waveforms using each possible subset of points while matching the initial time and velocity. Since modulating a waveform has a relatively low computational cost, we can explore each subset of points using a high-performance computing cluster. We calculate our residual through the relative difference in the systolic and diastolic area under the curve \cite{hozumi1998noninvasive}, as well as the relative two-norm difference ($r_2$), 

%We repeat this for five separate waveforms: four chosen to differ in shape by eye, and one that is the closest of the 4.5 million to the average. 

$$r_2 = \sqrt{\left(\frac{||t_w-t_g||_2}{||t_w||_2}\right)^2+ \left(\frac{||w-g||_2}{||w||_2}\right)^2}.$$

%------------------ RESULTS ------------------
\section{Results}
%To assess the results, we start by establishing the rationale for the stretching operation through an analysis of the correlation among various points. Following this, we calculate the residual for every conceivable combination of points. Finally, we delve into identifying and analyzing the optimal sets of points, considering each possible size of these sets.

% AR: Our results demonstrate the effectiveness of the stretching operation by analyzing the correlation among various points and computing the residual for all possible point combinations. We then identify the optimal sets of points for different set sizes, providing insights into the most efficient ways to quantify the velocity waveform.

Our results demonstrate the effectiveness of the stretching operation by computing the residual for all possible point combinations. We then identify the optimal sets of points for different set sizes, providing insights into the most efficient ways to quantify the velocity waveform.

\subsection{Selecting the points of interest from the fitting procedure}
\begin{figure}[tbp]
    \centering
    \includegraphics[width = \columnwidth]{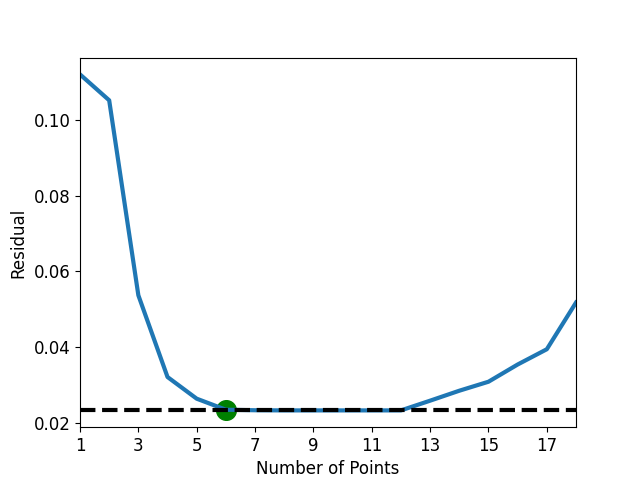}
    \caption{The smallest average residual for each point set size. The dashed horizontal line shows the minimal residual and the green dot denotes the smallest set size that can obtain the minimal residual.}
    \label{fig: residual}
\end{figure}

\begin{figure}[tbp]
    \centering
    \includegraphics[width = \columnwidth]{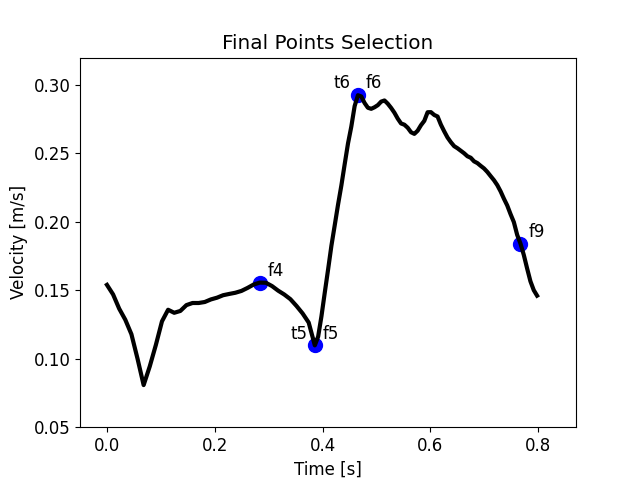}
    \caption{Points of interest that obtain the lowest residual for set size 6 (f4, t5, f5, t6, f6, f9), which corresponds to the green dot in Fig. \ref{fig: residual}.}
    \label{fig: best points on waves}
\end{figure}

To find the best set of points to characterize the waveform of the left coronary artery, we apply our stretching technique to fit a base waveform to the 1,000 representative waveforms chosen. We fitted using every possible set of points and computed the average residual across the 1,000 waveforms. Fig. \ref{fig: residual} shows the minimal residual for each set size. For example, of all sets with two points, the set with the smallest residual had a residual of just over 0.1. We observe that set sizes 6 through 12 obtained the same minimal residual. Furthermore, the points of interest that produce the smallest residual for the set size six can be seen in Fig. \ref{fig: best points on waves}.

% \begin{figure}[htbp]
% \centerline{\includegraphics{fig1.png}}
% \caption{Example of a figure caption.}
% \label{fig}
% \end{figure}

%------------------ DISCUSSION ------------------
\section{Discussion}

We have developed a pipeline that finds the minimum number of points of interest to quantify a velocity waveform and specifically consider left coronary artery velocity waveforms as proof of concept. We first verified that the proposed points of interest are positively correlated within the time and velocity domains. Once this correlation has been verified, we applied the fitting procedure to the waveforms in question and examined residuals to determine the most efficient way to quantify the velocity waveform.

\subsection{Positive Correlations of Time and Velocity Points}
The stretching transformation assumes the positive correlation of points. This assumption is most obvious in the time domain. For example, since $t_i < t_{i+1}$, if $t_i$ increases then $t_{i+1}$ is likely to increase as well. The velocity domain is less obvious, but one can infer that if a velocity point increases, then the rest of the velocity points are also likely to increase as a result of the inertia of blood. Thus, for blood flow velocity waveforms, the stretching procedure is well-founded.

\subsection{Selecting the Optimal Points to Characterize a Blood Velocity Waveform}
We base our selection of points of interest on the assumption that the points that can best be used to stretch a base waveform to the ones of interest are the most important to the shape of waveforms. This assumption is motivated by the ability of these points to change the shape of the base waveform.

For left coronary arteries we see from Fig. \ref{fig: residual} that the minimum residual is present for point set sizes of 6 to 12. In this case, the first set size that achieves the minimum residual is a set with 6 points, however; the difference in residual between a set of 5 points and 6 is minimal, so a set of either 5 or 6 points is both appropriate.

\subsection{Applications in the Clinic}
Finding the minimum number of points that adequately characterize a waveform is of clinical significance as it will allow clinicians to effectively quantitatively assess waveforms. Previous studies \cite{hozumi1998noninvasive,ofili1993coronary} have demonstrated the importance of the area under the systolic and diastolic curves, as well as the maximum velocity, but are coarse metrics. The method presented here allows clinicians to focus on a small number of observable points to characterize the waveform.

% However, these metrics are only coarse measurements that do not take into account the intricacies of the waveform shape. Additionally, examining all potentially interesting points, such as the 20 original points, would produce an overwhelming amount of information and redundancy that could potentially slow or inhibit clinical practice.

% Assessing the impacts of boundary conditions in computational fluid dynamics (CFD) has been the source of a plethora of studies. A variety of methods have been used to estimate the inlet temporal profile of blood flow, ranging from using a waveform from literature \cite{vardhan2019importance} or using 0D models to estimate the profile \cite{grande2021computational}. However, quantification of the effects of different waveforms has remained difficult because of the field's inability to characterize a waveform. The results of this study will allow CFD scientists to purposefully modulate selected points of interest to quantify their effects.

\subsection{Limitations}
In this proof-of-concept study, we used a database of simulated coronary artery waveforms \cite{tanade2023cloud}. Future work should apply this pipeline to patient waveforms. We do not label the bump in diastole with a point, as it is difficult to find automatically and preliminary analysis showed that it is minimally relevant to the residual.

% These waveforms are generated in part by scaling the time and the velocity which results in strong positive correlations between points. However, we also motivate this correlation physiologically and f

\subsection{Conclusion}
%We have presented a pipeline that, given waveforms and the identification of the maximum number of important features, can produce the smallest set of points that can account for the shape of a waveform. Additionally, we applied this pipeline to simulated left coronary artery waveforms as a proof of concept and found that a relatively small number of points can be used to recover the variations in waveforms. Future studies should employ this pipeline to other physiological waveforms.

In our study, we introduced a method that efficiently identifies and utilizes the minimal number of significant features from given waveforms. This approach ensures the accurate representation of a waveform's shape with a reduced set of points. As a practical demonstration, we applied this method to simulated left coronary artery waveforms and discovered that a surprisingly small selection of points suffices to capture waveform variations. This finding opens the door to future research where this pipeline could be applied to a variety of other physiological waveforms for similar efficiency and effectiveness.

% \section*{Acknowledgment}

\printbibliography

\end{document}